\documentclass[mathleft
]{an}
\usepackage{graphicx}
\usepackage{times}
\begin{document}

\Pagespan{789}{}
\Yearpublication{2012}%
\Yearsubmission{2012}%
\Month{00}%
\Volume{999}%
\Issue{88}%

\title{X-ray measurement of the elemental abundances at the outskirts of the Perseus cluster with \textit{Suzaku}}

\author{S.~Ueda\inst{1}\fnmsep\thanks{Corresponding author:
  \email{shutaro@ess.sci.osaka-u.ac.jp}\newline}
, K.~Hayashida\inst{1}, H.~Nakajima\inst{1}, 
\and  H.~Tsunemi\inst{1}
}
\titlerunning{Elemental abundances at the outskirts of the Perseus cluster}
\authorrunning{S.~Ueda et al}
\institute{
Department of Earth and Space Science, Graduate School of Science, Osaka University, Toyonaka, Osaka, 560-0043, Japan
}

\received{24 August 2012}
\accepted{16 October 2012}
\publonline{later}

\keywords{galaxies: clusters: individual (Perseus cluster) -- X-rays: galaxies: clusters}

\abstract{%
We report on the abundance of metals (Mg and Fe) 
in the intracluster medium (ICM) at the outskirts (0.2\,$r_{200} -$ 0.8\,$r_{200}$) of the Perseus cluster.
The X-ray spectra were obtained in the \textit{Suzaku}/XIS mapping observations of this region.
We employ single temperature models to fit all the X-ray spectra.   
The ICM temperature smoothly decreases toward the outer region from 6\,keV to 4\,keV.
The Fe abundance is uniformly distributed at the outskirts ($\sim0.3$\,solar). 
The Mg abundance is $\sim$1\,solar at the outskirts.
The solar ratios of Mg/Fe of the outskirts region (Mg/Fe $\sim$4) are a factor of 4 larger than those of the central region.
Various systematic effects, including the spatial fluctuations in the cosmic X-ray background, are taken into account and evaluated.
These our results have not changed significantly.
}

\maketitle

\section{Introduction}

Metals in the intracluster medium (ICM) was discovered by \cite{Mitchell} who detected He-like Fe lines in the Perseus cluster for the first time.
The metal abundance in the ICM has been measured in a number of clusters with its spatial distribution.
\cite{Matsushita} reported that Fe abundance distribution is centrally peaked in clusters with a cD galaxy,
indicating a significant amount of metals at cluster cores is supplied from those cD galaxies.
Recent X-ray observations extended the area of the metallicity measurement to and beyond virial radii (e.g. \cite{Fujita}, \cite{Simionescu}, and \cite{Urban});
metals are distributed at the virial radii of clusters with typical abundance (primarily determined from Fe) of 0.3\,solar.  
These results suggest that the ICM is not a pure pristine gas but polluted by metals from galaxies even in the outskirts of clusters.

Significant fraction of metals in the ICM are generated by supernovae.
Type Ia supernovae (SNe Ia) produce a large amount of Fe, while Type II supernovae (SNe II) synthesize lighter elements than Fe such as O, Ne, and Mg.
The abundance of each light element (O, Ne, and Mg) in the ICM was measured in the core of some bright clusters (e.g. \cite{Sato}, Tamura et al. 2009, and \cite{Sakuma}).
On the other hand, measurements of the metal abundances at the cluster outskirts have been limited to Fe. 
Therefore contribution of each type of SNe is difficult to be evaluated for them. 

\textit{Suzaku}, the 5th Japanese X-ray astronomy satellite, carries the X-ray Imaging Spectrometer (XIS: \cite{Koyama}) consisting of four X-ray CCD cameras.
Advantages of the XIS include its large effective area, low non X-ray background (NXB), and good reproducibility of the NXB.
Therefore, the XIS is a powerful tool for observing faint and diffuse objects such as the outskirts of clusters as already demonstrated (e.g. \cite{Hoshino}, \cite{Simionescu}).

The Perseus cluster, as the X-ray brightest and one of the nearest clusters, is the best target to measure the spatial distribution of various elements in the ICM.
\cite{Dupke} measured the Fe abundance at the outer region ($25^{\prime}-45^{\prime}$) using \textit{ASCA}.
\cite{Tamura} discovered rare elements (Cr and Mn) in the core using \textit{Suzaku}/XIS.
Simionescu et al. (2011) reported the large scale structure of the ICM beyond virial radius with \textit{Suzaku}/XIS.
They reported that the metallicity of the ICM at the outskirts is $\sim$0.3\,solar throughout the virial radius, 
although it is measured commonly for all the elements, predominantly determined by the Fe abundance.  
Mapping observations of the Perseus cluster have been extensively performed, and the area covered is several times larger than those used in Simionescu et al. (2011).
We focus on the Mg and Fe abundances of the ICM at the outskirts of the Perseus cluster ($0.2\,r_{200}  - 0.8\,r_{200}$).

In this paper, we adopt the abundance table of \cite{Anders} and the Hubble constant of $H_{0} = 70$\,km s$^{-1}$ Mpc$^{-1}$.
1\,arcmin corresponds to 22\,kpc at the redshift of this cluster, i.e., 0.0183, and 
the virial radius ($r_{200}$) is 1.79\,Mpc (\cite{Simionescu}).
Unless otherwise specified, all errors represent at 90\% confidence level (90$\%$CL).

\section{Observations and data reduction}

\begin{figure*}
	\begin{center}
		\includegraphics[width=130mm,height=110mm]{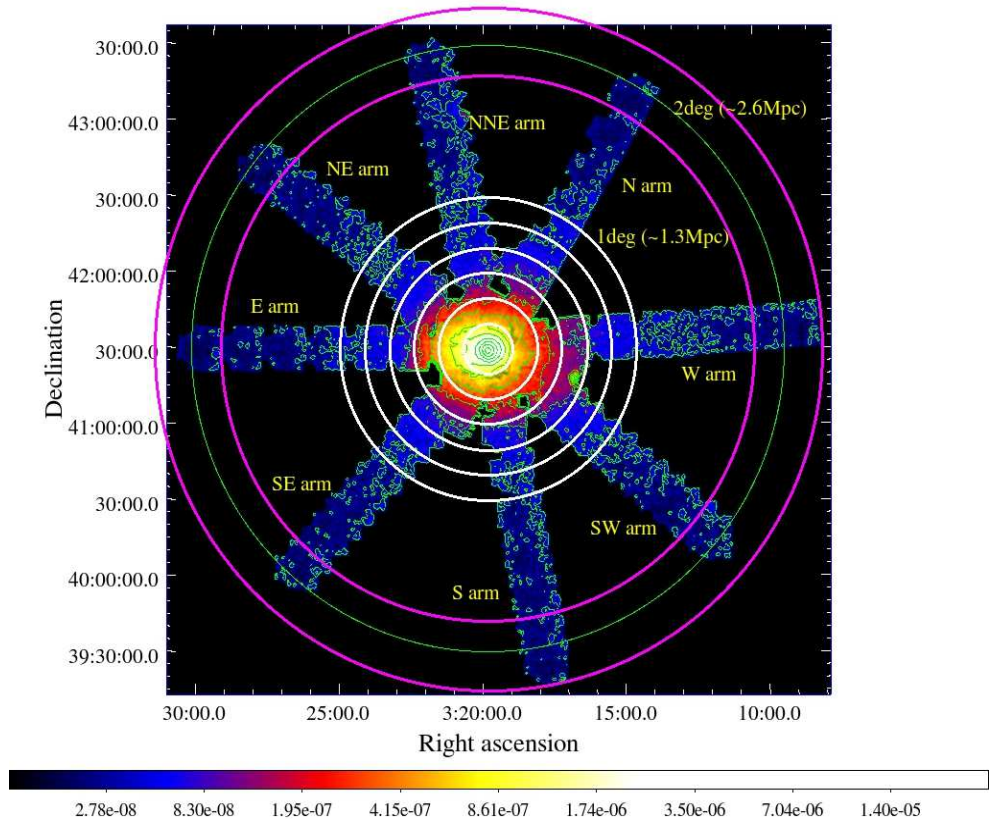}
	\end{center}
\caption{
X-ray image of the Perseus cluster.
Vignetting and exposure corrections were applied. The NXB was not subtracted, though its contribution is smaller than 30\% of the total flux.
The unit of color bar is counts s$^{-1}$ pixel$^{-1}$.
X-ray spectra were extracted from 5 annular regions bordered with white circles (their radius ranges from 10$^{\prime}$ to 60$^{\prime}$ $\sim$1.3\,Mpc).
The CXB and the Galactic emissions were estimated from the X-ray spectra extracted from the regions in the magenta annulus, 
whose distance is about 2.6\,Mpc from the center.
}
\label{fig:Perseus}
\end{figure*}

\textit{Suzaku} mapping observations of the Perseus cluster were performed from July 2009 to September 2011 (PI: S. W. Allen).
These observations form 8 arms, each of which consists of 7 - 12 pointings.
Fig.~\ref{fig:Perseus} shows the X-ray image of the Perseus cluster.
Only the observations of the N and E arms in Fig.~\ref{fig:Perseus} were employed in Simionescu et al. (2011).
Although each arm extends to $2^{\circ}$ (2.6\,Mpc or 1.6\,$r_{200}$) from the center,
we extracted the X-ray spectra from the region of $10^{\prime} \le r \le 60^{\prime}$, dividing it into 5 annuli in steps of 10$^{\prime}$,
as shown in Fig.~\ref{fig:Perseus} with white circles.
The outermost radius of 60$^{\prime}$ corresponds to 0.8\,$r_{200}$.
We excluded a circular region with radius of 5$^{\prime}$ around the the bright X-ray source IC 310.

\section{Spectral analyses}
\label{sec:spec}

We employed the emission model of thin thermal plasma with a single temperature (vAPEC, \cite{Smith}) to reproduce the X-ray spectra of the ICM.
The NXB was estimated by using the database of night earth observations with \texttt{xisnxbgen} (\cite{Tawa}) and subtracted.
The cosmic X-ray background (CXB) and the Galactic emissions were included as spectral models.

\subsection{Components of the CXB and the Galactic emissions}
\label{sec:cxb}

The CXB and the Galactic emissions (MWH: Milky Way halo and LHB: local hot bubble) were reproduced with a cut-off power-law (PL) and a two-component APEC model, respectively.
The schematic model was described as wabs $\times$ (vAPEC $+$ cut-off PL $+$ APEC$_{\rm MWH}$) $+$ APEC$_{\rm LHB}$, 
where wabs model represented the Galactic absorption using Wisconsin cross-sections (\cite{Morrison}).
We fixed $N_{\rm H}$ at $0.132 \times 10^{22}$ cm$^{-2}$, which was determined in the X-ray observations of the central region of this cluster (Nishino et al. 2010).
This value is consistent with that from 21-cm observations (Kalberla et al. 2005).

We determined the CXB and the Galactic emissions
by fitting the X-ray spectra of the outermost regions beyond 2.3\,Mpc (as shown with magenta annulus in Fig.~\ref{fig:Perseus}).
The cut-off energy of the cut-off PL for the CXB was fixed to 40\,keV, according to \cite{Boldt}.
The photon index and the flux of cut-off PL are 1.43$\pm$0.03 and 2.37$\pm 0.08\times10^{-11}$ ergs cm$^{-2}$ s$^{-1}$ deg$^{-2}$ ($2-10$\,keV), respectively.
These are consistent with those reported by \cite{Moretti}.
For the Galactic emissions, the abundance was set to 1\,solar for both models.
The gas temperatures are 0.23$\pm$0.02\,keV (MWH) and 0.07$\pm$ 0.02\,keV (LHB), respectively. 

\subsection{ICM component}
\label{sec:ICMcom}

\begin{figure}
  \begin{center}
    \includegraphics[width=70mm,height=50mm]{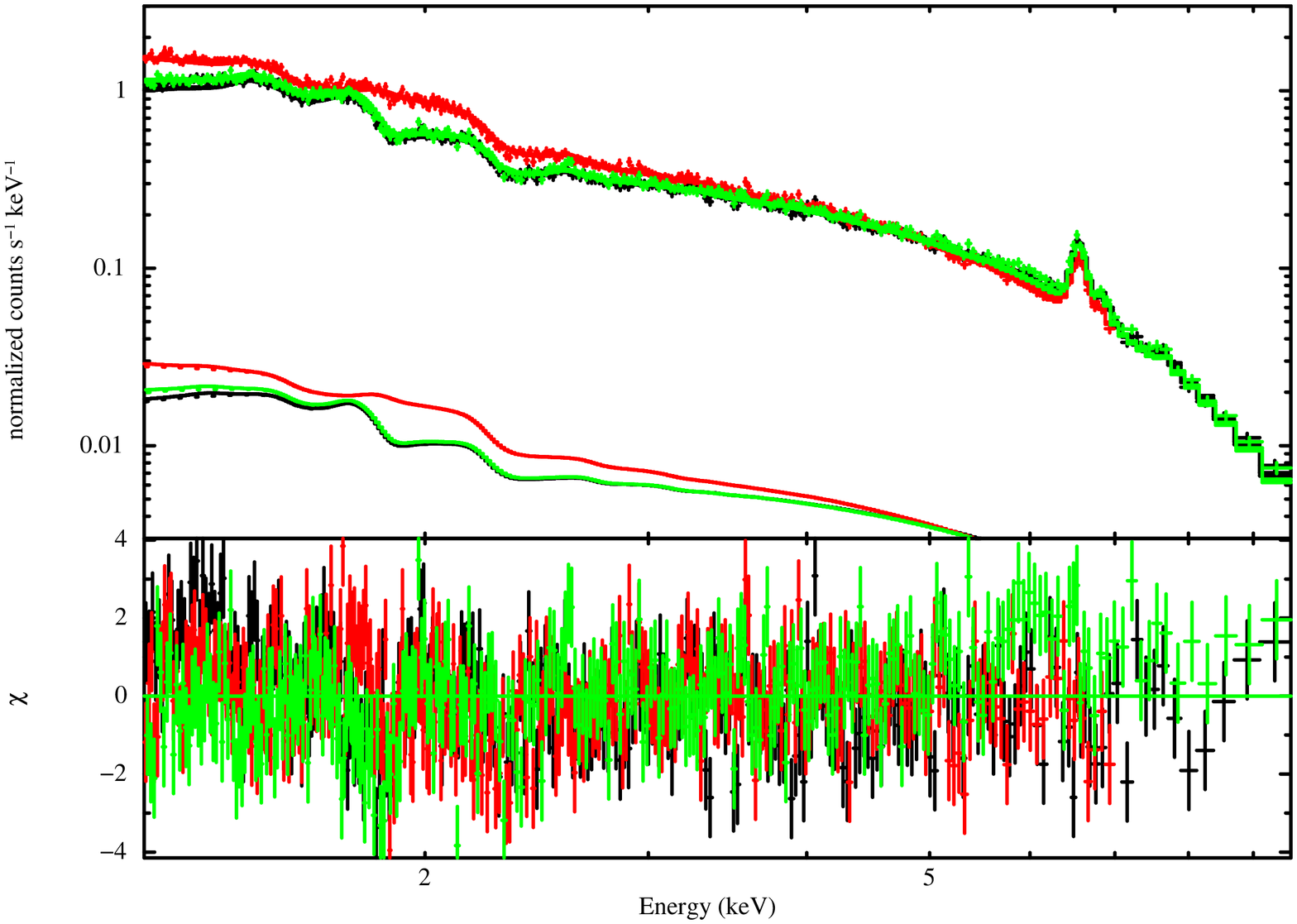}
    \includegraphics[width=70mm,height=50mm]{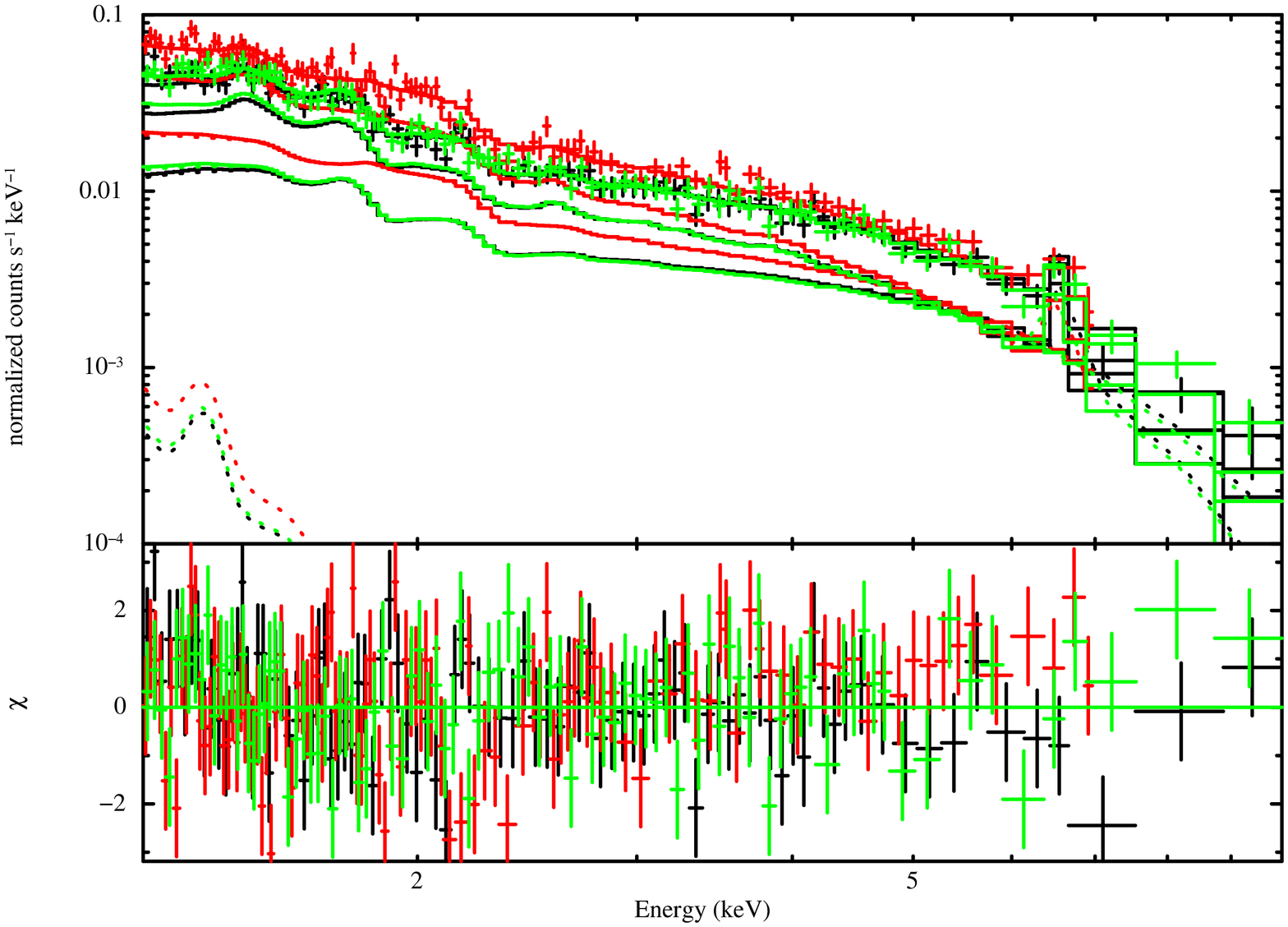}
  \end{center}
  \caption{
Top: X-ray spectra of the innermost region of our analyses ($10^{\prime} - 20^{\prime}$).
Single temperature model (vAPEC) is used for the ICM components in the fit. 
The CXB and the Galactic emissions were modeled as mentioned in \ref{sec:cxb}.
The spectra are plotted by black, red, and green across for those of XIS0, XIS1, and XIS3, respectively.
Bottom: Same as the top panel but for the outermost region ($50^{\prime} - 60^{\prime}$).
}
\label{fig:spectra}
\end{figure}

For the ICM (vAPEC) component, the abundances of He, C, N, O, Ne, and Al were fixed to 1\,solar, and Ni was linked to Fe.
We fitted the X-ray spectra in $1.2-10.0$\,keV at XIS0, 3 and $1.2-7.0$\,keV at XIS1 to avoid uncertainties of N$_{\rm H}$ and the contamination on OBF (\cite{Koyama}).
We focus on Mg and Fe abundances in this paper.
The top panel of Fig.~\ref{fig:spectra} shows the X-ray spectra at the innermost region ($10^{\prime} - 20^{\prime}$).
The bottom panel of Fig.~\ref{fig:spectra} shows those of the outermost region ($50^{\prime} - 60^{\prime}$).
The $\chi^2$/d.o.f. of the innermost region is 1243/851 and that of the outermost region is 460/389.

\section{Results}

\begin{figure*}
  \begin{center}
    \includegraphics[width=56mm,height=43mm]{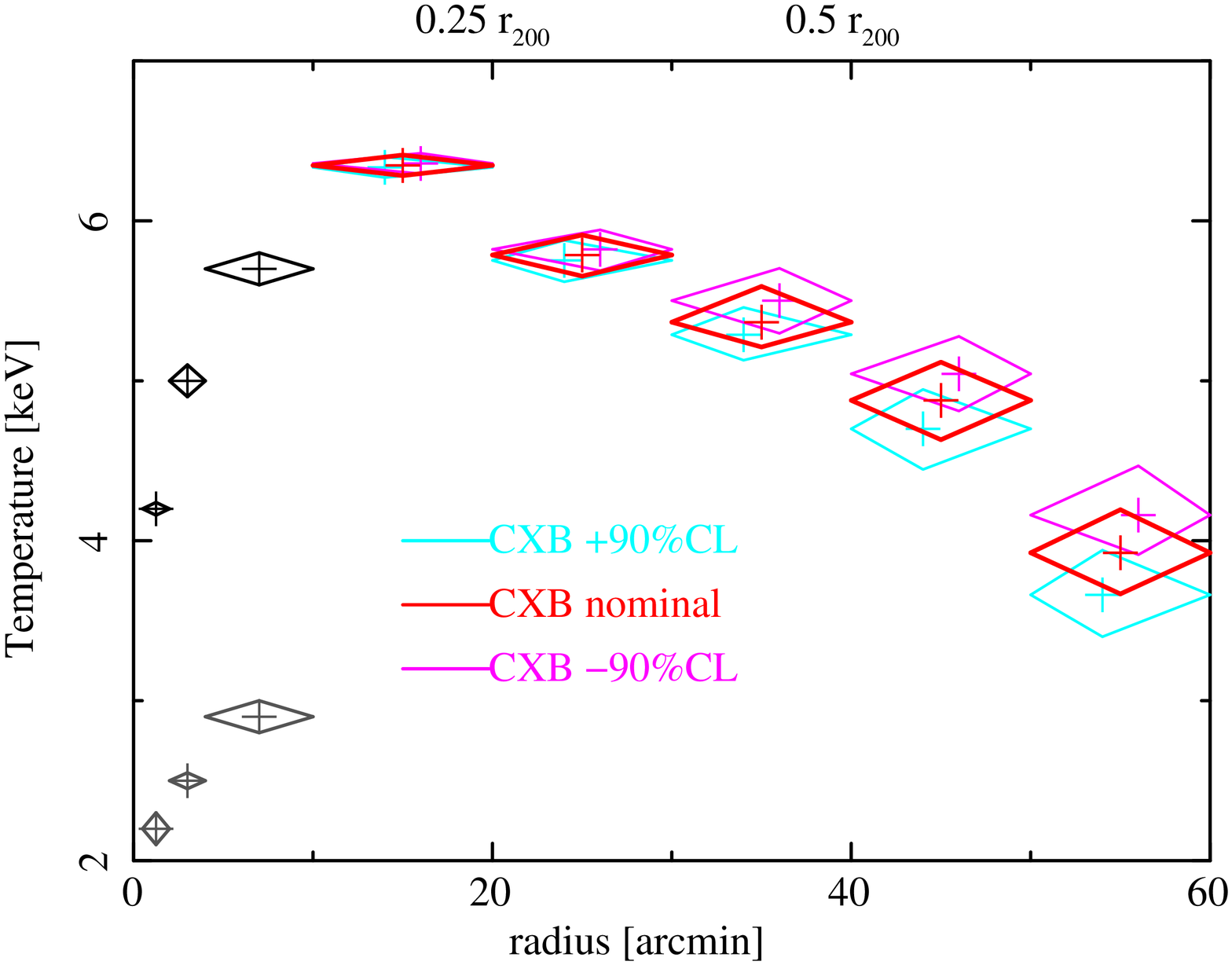}
    \includegraphics[width=56mm,height=43mm]{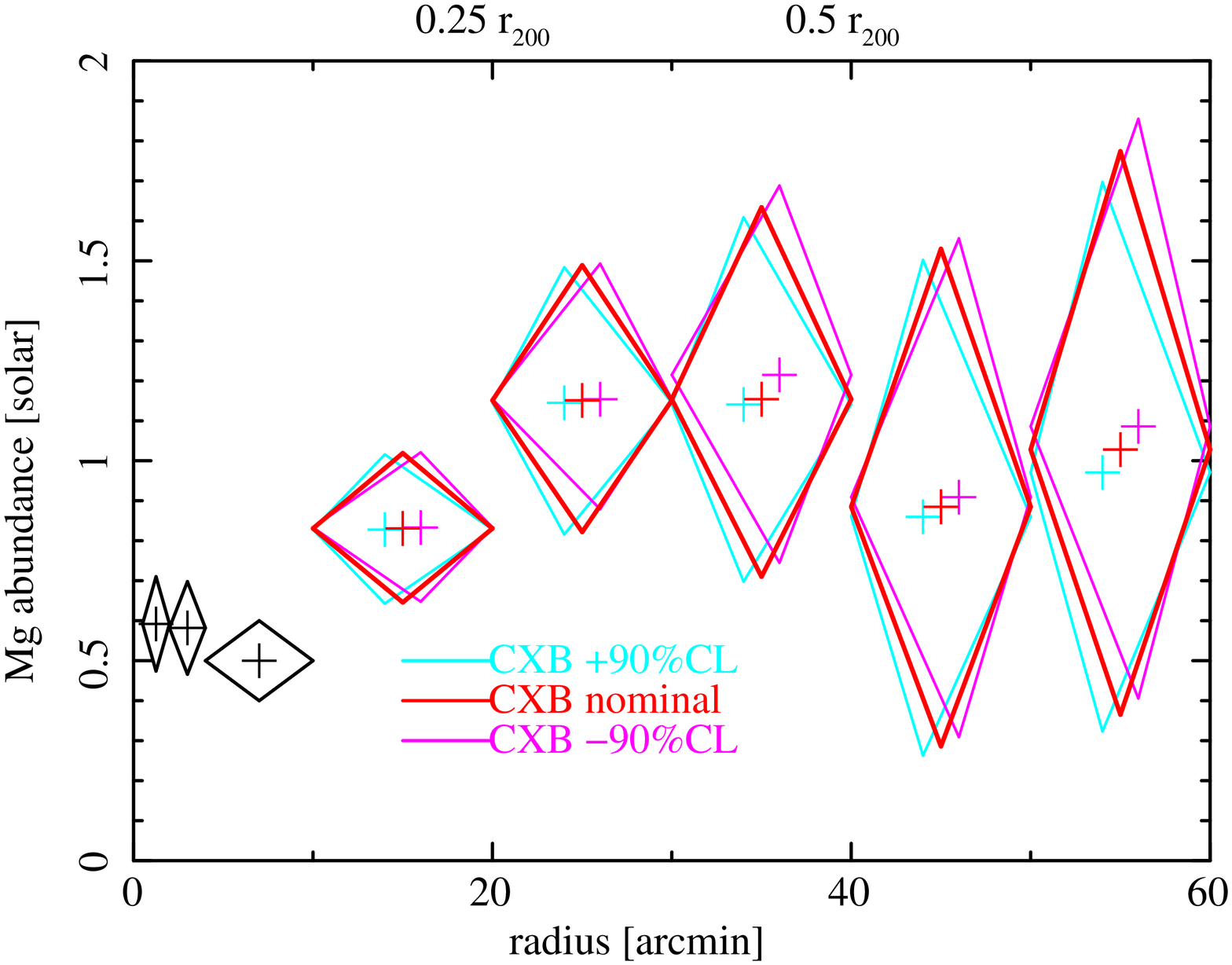}
    \includegraphics[width=56mm,height=43mm]{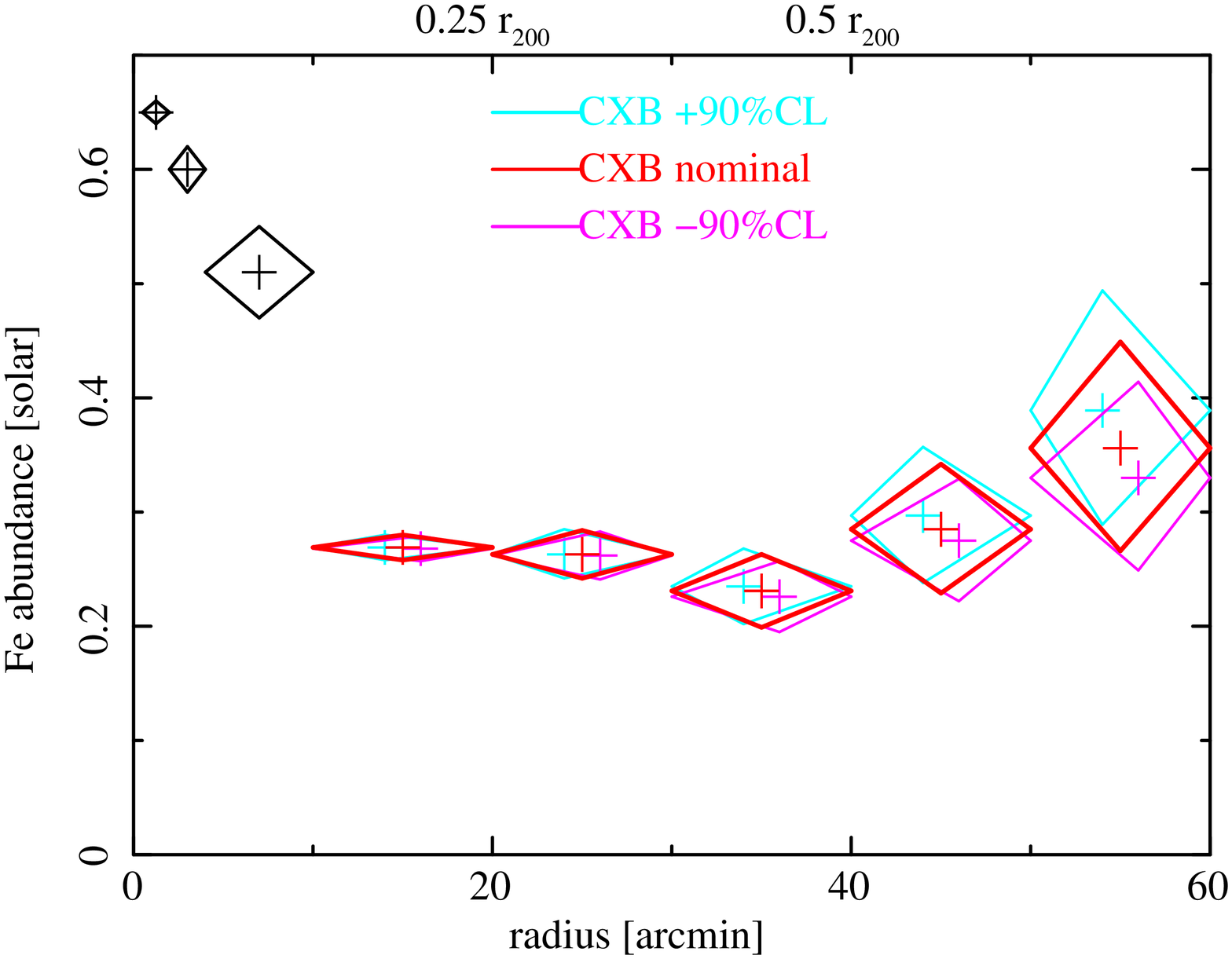}
  \end{center}
\caption{
Left: Radial profile of the ICM temperature including systematic errors ($\pm90\%$CL) owing to the spatial fluctuation of the CXB.
Middle: Same as the left panel but for the Mg abundance.
Right: Same as the central panel but for the Fe abundance.
The black diamonds in the inner regions ($\le10^{\prime}$) are taken from \cite{Tamura}.
The red diamonds show the temperature, the Mg and Fe abundances of the ICM in each annular region with the nominal flux of the CXB.
The light blue and magenta diamonds show the results when we changed the $+90\%$CL or $-90\%$CL of the CXB flux, respectively.
}
	\label{fig:CXBfluc}
\end{figure*}

The radial profile of the ICM temperature is shown in the left panel of Fig.~\ref{fig:CXBfluc}.
The ICM temperatures of the central regions ($\le 10^{\prime}$) are taken from \cite{Tamura}.
The ICM temperature smoothly decreases from 6\,keV to 4\,keV toward the outside.
This result is consistent with that obtained by Simionescu et al. (2011).

\begin{figure}
  \begin{center}
    \includegraphics[width=80mm,height=60mm]{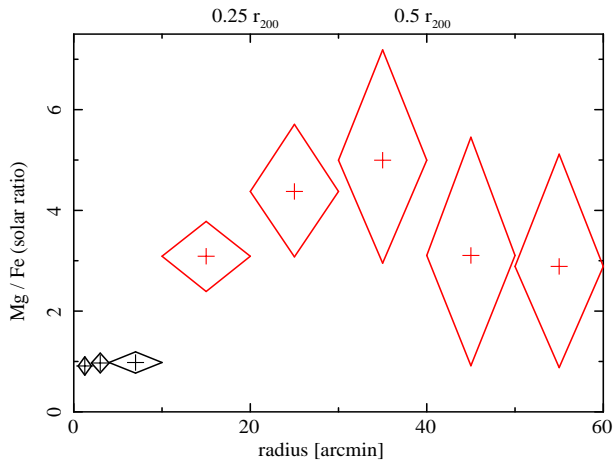}
  \end{center}
\caption{
Radial profile of the solar ratio of the Mg abundance to the Fe abundance.
}
\label{fig:Mg_Fe}
\end{figure}

We measured the Mg and Fe abundances in $0.2\,r_{200} - 0.8\,r_{200}$ region.
The radial profiles of the Mg and Fe abundances are shown in the middle panel and the right panel of Fig.~\ref{fig:CXBfluc}, respectively.
The Fe abundance is uniformly distributed at the outskirts ($\sim0.3$\,solar).
The Mg abundance is $\sim$1\,solar in the same region. 
We take the solar ratio of Mg abundance to that of the Fe (Mg/Fe ratio) in order to minimize the systematic errors to uncertainties in the plasma model for Fe-L complex. 
The result is shown in Fig.~\ref{fig:Mg_Fe}.
The mean values of Mg/Fe ratios of the central regions is $\displaystyle 0.95\pm0.11$, 
while that of the outskirts is $\displaystyle 3.7\pm0.8$.
We carried out the $t$-test in order to verify the statistical difference between these two value.
These two values are statistically different, i.e., the hypothesis that these two values are the same is rejected at a significance level of 0.001.

We evaluated the systematic errors owing to the spatial fluctuation of the CXB flux in our results.
Tawa et al. (2008) estimated that the fluctuation in the CXB flux is 14\% (1$\sigma$) for the whole area of the XIS.
Assuming the fluctuation has been inversely proportional to the root mean square of the size of the sky area, 
the magnitude of the fluctuation for each annular region was calculated.
We added and subtracted the 90$\%$CL of the fluctuation to the CXB nominal flux model (as mentioned in \ref{sec:cxb}) 
when we fitted the X-ray spectra of in each annular region.
As shown in Fig.~\ref{fig:CXBfluc} (light blue diamonds for $+90\%$CL, those of magenta for $-90\%$CL),
the systematic errors due to the CXB fluctuations are much smaller than the statistical errors in the radial profiles of the ICM temperature, the Mg, and Fe abundance. 

We also examined the systematic errors due to fixed the abundances of C, N, O, Ne, and Al on the abundances of Mg and Fe.
We fixed the abundances of these elements at 1\,solar.
We tested the spectral analysis of $10^{\prime} - 20^{\prime}$ region by changing the abundances of these elements to 0.5\,solar and 2\,solar. 
This changes the best fit values of the Mg and Fe abundance by at most 13\%. 
We also checked the possibility that the Al abundance affects the Mg abundance. 
Even if we take the extreme case, i.e., 0.1\,solar and 10\,solar for the Al abundance, the Mg abundance differs by 15\%.
These systematic effect cannot explain the factor of 4 difference in the Mg/Fe ratios we observed. 
Note that although we fixed the He abundance at 1\,solar, as has been commonly assumed, changing this value scales all abundances of elements heavier than C (e.g. \cite{Ettori}).  

Another concern is the difference in the emission model we assumed, i.e., single temperature or two temperatures. 
Nevertheless, it is found that the Mg abundance in the central ($4^{\prime}-10^{\prime}$ region) is altered by at most $+21\%$ 
if we fit the spectra with a single temperature model, instead of the two-temperature model.

The Fe abundance in the ICM is centrally peaked, while that of the outskirts is almost constant.
This result suggests that significant amount of Fe is supplied from the cD galaxy where the contribution of SNe Ia is relatively large.
Larger Mg/Fe ratio at the outskirts than that in the central region is another support on this point of view.

\section{Summary}

We summarize our results on the ICM at the outskirts of the Perseus cluster as follows:

\begin{enumerate}

\item We measured the metal abundances (Mg and Fe) separately at $0.2\,r_{200} - 0.8\,r_{200}$.

\item The Fe abundance is constant of $\sim$0.3\,solar at the outskirts.

\item The Mg abundance is $\sim$1\,solar in the outskirts region. 

\item The mean value of Mg/Fe ratio is $\displaystyle 0.95\pm0.11$ in the central region, while that in the outskirts is $\displaystyle 3.7\pm0.8$. 
This difference is significant even we consider the possible systematic effects. 
Centrally peaked Fe abundance is likely due to large contribution of SNe Ia products from the cD galaxy. 
The difference in the Mg/Fe ratio we observe is consistent with this point of view.

\end{enumerate}

\acknowledgements
We are grateful to the anonymous referee for helpful suggestions and comments.
We thank all members of the \textit{Suzaku} operation team and the XIS calibration team.
We also thank A. Simionescu, N. Werner, Y. Takei, and T. Tamura for their useful comments.
SU is supported by Japan Society for the Promotion of Science (JSPS) Research Fellowship for Young Scientist (A2411900).
This work was also supported by JSPS KAKENHI Grand Number 23340071 (KH), 24684010 (HN), and 23000004 (HT).
 




\begin{thebibliography}{}

\bibitem [Anders \& Grevesse (1989)]{Anders}
  Anders,~E. \& Grevesse,~N.: 1989, Geochim. Cosmochim. Acta, 53, 197
  
\bibitem[Boldt (1987)]{Boldt}
  Boldt,~E.: 1987, Phys. Rep., 146, 215
  
\bibitem[Dupke \& Arnaud (2001)]{Dupke}
  Dupke,~R.~A., \& Arnaud,~K.~A.: 2001, \apj, 548, 141
  
\bibitem[Ettori \& Fabian 2006]{Ettori}
  Ettori, S., \& Fabian, A.~C.: 2006, \mnras, 369, L42
  
\bibitem[Fujita et al. 2008]{Fujita}
  Fujita,~Y., et al.: 2008, \pasj, 60, 343
  
\bibitem[Hoshino et al. 2010]{Hoshino}
  Hoshino, A., et al.: 2010, \pasj, 62, 371
    
\bibitem[Kalberla et al. 2005]{Kalberla} 
  Kalberla,~P.~M.~W., et al.: 2005, \aaa, 440, 775
    
\bibitem[Koyama et al. 2007]{Koyama} 
  Koyama,~K., et al: 2007, \pasj, 59, 23
  
\bibitem[Matsushita (2011)]{Matsushita}
  Matsushita,~K.: 2011, \aaa, 527, A134

\bibitem[Moretti et al. (2009)]{Moretti}
  Moretti,~A., et al.: 2009, \aaa, 493, 501
  
\bibitem[Morrison \& McCammon 1983]{Morrison}
  Morrison, R.,~\& McCammon,~D.: 1983, \apj, 270, 119
  
\bibitem[Mitchell et al. (1976)]{Mitchell} 
  Mitchell, R.~J., et al.: 1976, \mnras, 175, 29P
    
\bibitem[Nishino et al. 2010]{Nishino1}
  Nishino,~S., et al.: 2010, \pasj, 62, 9

\bibitem[Sakuma et al. 2011]{Sakuma} 
  Sakuma,~E., et al.: 2011, \pasj, 63, 979
  
\bibitem[Sato et al. 2008]{Sato}
  Sato,~K., et al.: 2008, \pasj, 60, S333
  
\bibitem[Simionescu et al. 2011]{Simionescu}
  Simionescu,~A., et al.: 2011, Science, 331, 1576
  
\bibitem[Smith et al. 2001]{Smith}
  Smith,~R.~K., et al.: 2001, \apj, 556, L91

\bibitem[Tamura et al. (2009)]{Tamura}
  Tamura,~T., et al.: 2009, \apj, 705, L62
  
\bibitem[Tawa et al. 2008]{Tawa}
  Tawa, N., et al.: \ 2008, \pasj, 60, S11
  
\bibitem[Urban et al. 2011]{Urban}
  Urban,~O., et al.: 2011, \mnras, 414, 2101


\end{thebibliography}
\end{document}